\documentclass[10pt]{article}
\usepackage{latexsym}
\usepackage{amssymb}
\usepackage{amsmath}
\usepackage{amscd}
\usepackage{amsthm}
\usepackage[left=2cm,top=2.5cm,right=2.5cm,bottom=1.5cm]{geometry}
\usepackage[dvips]{graphicx}
\usepackage{hyperref}
\begin{document}
\begin{center}
\large{\bf{ACCELERATING DARK ENERGY MODELS IN BIANCHI TYPE-V SPACE-TIME}} \\
\vspace{10mm} \normalsize{ANIRUDH PRADHAN}  \\
\vspace{5mm} \normalsize{\it{Department of Mathematics, Hindu Post-graduate College, Zamania-232 331, Ghazipur, India} \\
E-mail : {pradhan@iucaa.ernet.in, pradhan.anirudh@gmail.com}} \\
\vspace{10mm} \normalsize{HASSAN AMIRHASHCHI}  \\
\normalsize{\it{Department of Physics, Mahshahr Branch, Islamic Azad University, Mahshahr, Iran \\
E-mail: h.amirhashchi@mahshahriau.ac.ir}} \\
\end{center}
\vspace{10mm}
\begin{abstract}
Some new exact solutions of Einstein's field equations in a spatially homogeneous and anisotropic Bianchi type-V space-time 
with minimally interaction of perfect fluid and dark energy components have been obtained. To prevail the deterministic 
solution we choose the scale factor $a(t) = \sqrt{t^{n}e^{t}}$, which yields a time dependent deceleration parameter (DP), 
representing a model which generates a transition of the universe from the early decelerating phase to the recent accelerating 
phase. We find that for $n \geq 1$, the quintessence model is reproducible with present and expected future evolution of 
the universe. The other models (for $n < 1$), we observe the phantom scenario. The quintessence as well as phantom models 
approach to isotropy at late time. For different values of $n$, we can generate a class of physically viable DE models. The 
cosmic jerk parameter in our descended model is also found to be in good concordance with the recent data of astrophysical 
observations under appropriate condition. The physical and geometric properties of spatially homogeneous and anisotropic 
cosmological models are discussed.
\end{abstract}
\section{Introduction}
Recent observations of type Ia supernovae suggest that the expansion of the universe is accelerating and 
two-thirds of the total energy density exists in a dark energy component with negative pressure 
\cite{ref1,ref2}(for a recent review, see Padmanabhan \cite{ref3}; Copeland et al. \cite{ref4}). In addition, 
measurements of the cosmic microwave background \cite{ref5} and the galaxy power spectrum \cite{ref6} also 
indicate the existence of the dark energy. However, the observational data are far from being complete (for a recent 
review, see Perivolaropoulos \cite{ref7}; Jassal et al.\cite{ref8}). It is not even known what is the current value of the 
dark energy effective equation of state (EoS) parameter $\omega^{(de)} = p^{(de)}/\rho^{(de)}$ which lies close to 
$-1$: it could be equal to $-1$ (standard $\Lambda$CDM cosmology), a little bit upper than $-1$ (the quintessence 
dark energy) or less than $-1$ (phantom dark energy). While the possibility $\omega \ll -1$ is ruled out by current 
cosmological data from SN Ia (Supernovae Legacy Survey, Gold sample of Hubble Space Telescope) \cite{ref9,ref10}, 
CMB (WMAP, BOOMERANGE) \cite{ref11,ref12} and large scale structure (Sloan Digital Sky Survey) data\cite{ref13}, 
the dynamically evolving DE crossing the phantom divide line (PDL) ($\omega = -1$) is mildly favoured. The simplest 
candidate for the dark energy is a cosmological constant $\Lambda$, which has pressure $p^{(de)} = - \rho^{(de)}$. 
Specifically, a reliable model should explain why the present amount of the dark energy is so small compared with 
the fundamental scale (fine-tuning problem) and why it is comparable with the critical density today (coincidence problem) 
\cite{ref4}. That is why, the different forms of dynamically changing DE with an effective equation of state (EoS), 
$\omega^{(de)} = p^{(de)}/\rho^{(de)} < -1/3$, have been proposed in the literature. Some other limits obtained from 
observational results coming from SNe Ia data \cite{ref14} and combination of SNe Ia data with CMBR anisotropy and 
galaxy clustering statistics \cite{ref6} are $-1.67 < \omega < -0.62$ and $-1.33 < \omega < - 0.79$, respectively. The 
latest results in 2009, obtained after a combination of cosmological datasets coming from CMB anisotropies, luminosity 
distances of high redshift type Ia supernovae and galaxy clustering, constrain the dark energy EoS to 
$-1.44 < \omega < -0.92$ at $68\%$ confidence level \cite{ref15,ref13}. \\

Moreover, in recent years Bianchi universes have been gaining an increasing interest of observational cosmology, since 
the WMAP data \cite{ref15}$-$\cite{ref17} seem to require an addition to the standard cosmological model with positive 
cosmological constant that resembles the Bianchi morphology \cite{ref18}$-$\cite{ref23}. According to this, the universe 
should achieved a slightly anisotropic special geometry in spite of the inflation, contrary to generic inflationary 
models and that might be indicating a non-trivial isotropization history of universe due to the presence of an anisotropic 
energy source. In principle, once the metric is generalized to Bianchi types, the EoS parameter of the fluid can also be 
generalized in a way conveniently to wield anisotropy with the considered metric. In such models, where both the metric 
and EoS parameter of the fluid are allowed to exhibit an anisotropic character, the universe can exhibit non-trivial 
isotropization histories and it can be examined whether the metric and/or the EoS parameter of fluid evolve toward 
isotropy. Thus, the Bianchi models which remains anisotropic are of rather academical interest. The study of Bianchi 
type V cosmological models create more interest as these models contain isotropic special cases and permit arbitrary 
small anisotropy levels at some instant of cosmic time. This property makes them suitable as model of our universe. The 
homogeneous and isotropic Friedman-Robertson-Walker (FRW) cosmological models, which are used to describe standard 
cosmological models, are particular case of Bianchi type $I$, $V$ and $IX$ universes, according to whether the constant 
curvature of the physical three-space, $t$ = constant, is zero, negative or positive. \\

Recently, Yadav \cite{ref24} and Kumar \& Yadav \cite{ref25} have dealed with a spatially homogeneous and anisotropic 
Bianchi type-V DE models by considering constant DP. In this paper, we have considered minimally interacting perfect 
fluid and DE components with time dependent DP within the framework of totally anisotropic Bianchi-V models. This paper 
is organized as follows: the metric and the field equations are presented in Sect. $2$. Sect. $3$ deals with the exact 
solutions of the field equations and the physical behaviour of the model. Section $4$ deals with cosmic jerk parameter. 
Discussions and concluding remarks are given in Section $5$.
\section{The Metric and the Field Equations}
We consider the space time metric of the spatially homogeneous and anisotropic Bianchi type-V of the form
\begin{equation}
\label{eq1} ds^{2} = - dt^{2} + A^{2} dx^{2} + e^{2\alpha x} \left[B^{2}dy^{2} + C^{2}dz^{2}\right],
\end{equation}
where the metric potentials $\rm A$, $\rm B$ and $\rm C$ are functions of cosmic time $\rm t$ alone and $\rm \alpha$ 
is a constant. \\

We define the following physical and geometric parameters to be used in formulating the law and further in solving 
the Einstein's field equations for the metric (\ref{eq1}). \\

The average scale factor $a$ of Bianchi type-V model (\ref{eq1}) is defined as
\begin{equation}
\label{eq2} a = (ABC)^{\frac{1}{3}}.
\end{equation}
A volume scale factor V is given by
\begin{equation}
\label{eq3} V = a^{3} = ABC.
\end{equation}
We define the generalized mean Hubble's parameter $\rm H$ as
\begin{equation}
\label{eq4} H = \frac{1}{3}(H_{x} + H_{y} + H_{z}),
\end{equation}
where $\rm H_{x} = \frac{\dot{A}}{A}$, $\rm H_{y} = \frac{\dot{B}}{B}$ and $\rm H_{z} = \frac{\dot{C}}{C}$ are the
directional Hubble's parameters in the directions of $x$, $y$ and $z$ respectively. A dot stands for differentiation 
with respect to cosmic time $t$. \\

From Eqs. (\ref{eq2})-(\ref{eq4}), we obtain
\begin{equation}
\label{eq5} H = \frac{1}{3}\frac{\dot{V}}{V} = \frac{\dot{a}}{a} = \frac{1}{3}\left(\frac{\dot{A}}{A} + 
\frac{\dot{B}}{B} + \frac{\dot{C}}{C}\right).
\end{equation}
The physical quantities of observational interest in cosmology i.e. the expansion scalar $\theta$, the average 
anisotropy parameter $Am$ and the shear scalar $\sigma^{2}$ are defined as
\begin{equation}
\label{eq6} \theta = u^{i}_{;i} = \left(\frac{\dot{A}}{A} + \frac{\dot{B}}{B} + \frac{\dot{C}}{C} \right),
\end{equation}
\begin{equation}
\label{eq7} \sigma^{2} =  \frac{1}{2} \sigma_{ij}\sigma^{ij} = \frac{1}{2}\left[\frac{\dot{A}^{2}}{A^{2}} +
\frac{\dot{B}^{2}}{B^{2}} + \frac{\dot{C}^{2}}{C^{2}}\right] - \frac{\theta^{2}}{6},
\end{equation}
\begin{equation}
\label{eq8} A_{m} = \frac{1}{3}\sum_{i = 1}^{3}{\left(\frac{\triangle H_{i}}{H}\right)^{2}},
\end{equation}
where $\triangle H_{i} = H_{i} - H (i = x, y, z)$ represents the directional Hubble parameter in the direction of 
$x$, $y$, $z$ respectively. $A_{m} = 0$ corresponds to isotropic expansion. \\

The Einstein's field equations ( in gravitational units $8\pi G = c = 1 $) read as
\begin{equation}
\label{eq9} R^{i}_{j} - \frac{1}{2} R g^{i}_{j} = - T^{(m)i}_{j} - T^{(de)i}_{j},
\end{equation}
where $T^{(m)i}_{j}$ and $T^{(de)i}_{j}$ are the energy momentum tensors of perfect fluid and DE, respectively. These 
are given by 
\[
  T^{(m)i}_{j} = \mbox{diag}[-\rho^{(m)}, p^{(m)}, p^{(m)}, p^{(m)}],
\]
\begin{equation}
\label{eq10} ~ ~ ~ ~ ~ ~ ~ ~  = \mbox{diag}[-1, \omega^{(m)}, \omega^{(m)}, \omega^{(m)}]\rho^{m},
\end{equation}
and 
\[
 T^{(de)i}_{j} = \mbox{diag}[-\rho^{(de)}, p^{(de)}, p^{(de)}, p^{(de)}],
\]
\begin{equation}
\label{eq11} ~ ~ ~ ~ ~ ~ ~ ~ ~ ~ ~ ~ ~ ~ = \mbox{diag}[-1, \omega^{(de)}, \omega^{(de)}, \omega^{(de)}]\rho^{(de)},
\end{equation}
where $\rho^{(m)}$ and $p^{(m)}$ are, respectively the energy density and pressure of the perfect fluid component or 
ordinary baryonic matter while $\omega^{(m)} = p^{(m)}/\rho{(m)}$ is its EoS parameter. Similarly, $\rho^{(de)}$ and 
$p^{(de)}$ are, respectively the energy density and pressure of the DE component while $\omega^{(de)} = p^{(de)}/\rho^{(de)}$ 
is the corresponding EoS parameter. We assume the four velocity vector $u^{i} = (1, 0, 0, 0)$ satisfying $u^{i}u_{j} = -1$. \\

In a co-moving coordinate system ($u^{i} = \delta^{i}_{0}$), Einstein's field equations (\ref{eq9}) with (\ref{eq10}) 
and (\ref{eq11}) for B-V metric (\ref{eq1}) subsequently lead to the following system of equations: 
\begin{equation}
\label{eq12} \frac{\ddot{B}}{B} + \frac{\ddot{C}}{C} + \frac{\dot{B}\dot{C}}{BC} - \frac{\alpha^{2}} {A^{2}} 
= -\omega^{(m)}\rho^{(m)} - \omega^{(de)}\rho^{(de)},
\end{equation}
\begin{equation}
\label{eq13} \frac{\ddot{C}}{C} + \frac{\ddot{A}}{A} + \frac{\dot{C}\dot{A}}{CA} - \frac{\alpha^{2}} {A^{2}} 
= -\omega^{(m)}\rho^{(m)} - \omega^{(de)}\rho^{(de)},
\end{equation}
\begin{equation}
\label{eq14} \frac{\ddot{A}}{A} + \frac{\ddot{B}}{B} + \frac{\dot{A} \dot{B}}{AB} - \frac{\alpha^{2}} {A^{2}} 
= -\omega^{(m)}\rho^{(m)} - \omega^{(de)}\rho^{(de)},
\end{equation}
\begin{equation}
\label{eq15} \frac{\dot{A}\dot{B}}{AB} + \frac{\dot{A}\dot{C}}{AC} + \frac{\dot{B}\dot{C}}{BC} - \frac{3\alpha^{2}}{A^{2}} 
= \rho^{(m)} + \rho^{(de)},
\end{equation}
\begin{equation}
\label{eq16} \frac{2\dot{A}}{A} - \frac{\dot{B}}{B} - \frac{\dot{C}}{C} = 0.
\end{equation}
The law of energy-conservation equation ($T^{ij}_{;j} = 0$) yields
\begin{equation}
\label{eq17} \dot{\rho}^{(m)} + 3(1 + \omega^{(m)})\rho^{(m)}H + \dot{\rho}^{(de)} +3(1 + \omega^{(de)})\rho^{(de)}H = 0.
\end{equation}
The Raychaudhuri equation is found to be
\begin{equation}
\label{eq18} \dot{\theta} = - \left(1 + 3\omega^{(de)}\right)\rho^{(de)} - \frac{1}{3}\theta^{2} - 2\sigma^{2}.
\end{equation}
\section{Solution of the Field Equations and its Physical Significance}
In order to solve the field equations completely, we first assume that the perfect fluid and DE components interact 
minimally. Therefore, the energy momentum tensors of the two sources may be conserved separately. \\\\
The energy-conservation equation ($T^{(m)ij}_{;j} = 0$) of the perfect fluid gives
\begin{equation}
\label{eq19} \dot{\rho}^{(m)} + 3\rho^{(m)}(1 + \omega^{(m)})H  = 0,
\end{equation}
where as energy-conservation equation ($T^{(de)ij}_{;j} = 0$) of the DE component leads to
\begin{equation}
\label{eq20} \dot{\rho}^{(de)} + 3\rho^{(de)}(\omega^{(de)} + 1)H = 0.
\end{equation}
Following, Akarsu and Kilinc (2010), Yadav (2011) and Kumar \& Yadav (2011), we assume that the EoS parameter of the perfect 
fluid to be constant, that is,
\begin{equation}
\label{eq21}\omega^{(m)} = \frac{p^{(m)}}{\rho^{(m)}} = \mbox{const.} ,
\end{equation}
while $\omega^{(de)}$ has been permitted to be a function of time since the current cosmological data from SN Ia, CMB and 
large scale structures mildly favour dynamically evolving DE crossing the PDL as discussed in previous section. \\\\
Eq. (\ref{eq19}) can be integrated to obtain
\begin{equation}
\label{eq22} \rho^{(m)} = \rho_{0}a^{-3(\omega + 1)},
\end{equation}   
where $\rho_{0}$ is a positive constant of integration.\\\\
Integrating (\ref{eq16}) and engrossing the constant of integration in $B$ or $C$, without any loss of generality, we obtain
\begin{equation}
\label{eq23} A^{2} = B C. 
\end{equation}
We now use the technique following Kumar and Yadav \cite{ref25} to solve Einstein's field equations (\ref{eq12}) $-$ 
(\ref{eq15}). Subtracting (\ref{eq12}) from (\ref{eq13}), (\ref{eq12}) from (\ref{eq14}), and (\ref{eq13}) from 
(\ref{eq15}) and taking second integral of each, we obtain the following three relations respectively:
\begin{equation}
\label{eq24} \frac{A}{B} = d_{1}\exp{\left(k_{1}\int{\frac{dt}{a^{3}}}\right)},
\end{equation}
\begin{equation}
\label{eq25} \frac{A}{C} = d_{2}\exp{\left(k_{2}\int{\frac{dt}{a^{3}}}\right)},
\end{equation}
and
\begin{equation}
\label{eq26} \frac{B}{C} = d_{3}\exp{\left(k_{3}\int{\frac{dt}{a^{3}}}\right)},
\end{equation}
where $d_{1}$, $d_{2}$, $d_{3}$, $k_{1}$, $k_{2}$ and $k_{3}$ are constants of integration. From (\ref{eq24})$-$(\ref{eq26}) 
and (\ref{eq23}), the metric functions can be explicitly obtained as
\begin{equation}
\label{eq27} A(t) = a,
\end{equation}
\begin{equation}
\label{eq28} B(t) = m a\exp{\left(\ell \int{\frac{dt}{a^{3}}}\right)},
\end{equation}
\begin{equation}
\label{eq29} C(t) = \frac{a}{m}\exp{\left(-\ell \int{\frac{dt}{a^{3}}}\right)},
\end{equation}
where
\begin{equation}
\label{eq30} m = \sqrt[3]{(d_{2}d_{3})}, \; \; \ell = \frac{(k_{2} + k_{3})}{3}, \; \; d_{2} = d^{-1}_{1}, 
\; \; k_{2} = - k_{1}.
\end{equation}
Finally, following Saha et al. \cite{ref26}, we take following {\it ansatz} for the scale factor, where increase in term 
of time evolution is
\begin{equation}
\label{eq31} a(t) = \sqrt{t^{n}e^{t}},
\end{equation}
where $n$ is a positive constant. This {\it ansatz} generalized the one proposed by Amirhashchi et al. \cite{ref27}. In 
literature it is common to use a constant deceleration parameter \cite{ref24,ref25}, \cite{ref28}$-$\cite{ref32} as it 
duly gives a power law for metric function or corresponding quantity. The motivation to choose such time dependent DP 
is behind the fact that the universe is accelerated expansion at present as observed in recent observations of Type 
Ia supernova \cite{ref1,ref2,ref10,ref33,ref34} and CMB anisotropies \cite{ref35}$-$\cite{ref37} and decelerated 
expansion in the past. Also, the transition redshift from deceleration expansion to accelerated expansion is about 
$0.5$. Now for a Universe which was decelerating in past and accelerating at the present time, the DP must show 
signature flipping \cite{ref38}$-$\cite{ref40}. So, there is no scope for a constant DP at the present epoch. So, in general, 
the DP is not a constant but time variable. The motivation to choose such scale factor (\ref{eq31}) yields a time 
dependent DP. \\
\begin{figure}[ht]
\centering
\includegraphics[width=10cm,height=10cm,angle=0]{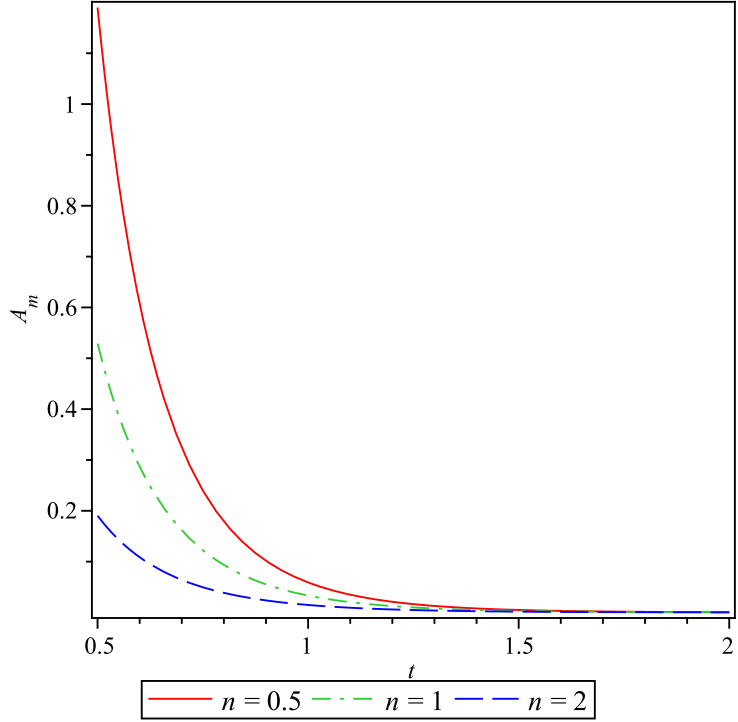} \\
\caption{The anisotropic parameter $A_{m}$ versus $t$. 
Here $\ell = 1$ }. 
\end{figure}
\begin{figure}[ht]
\centering
\includegraphics[width=10cm,height=10cm,angle=0]{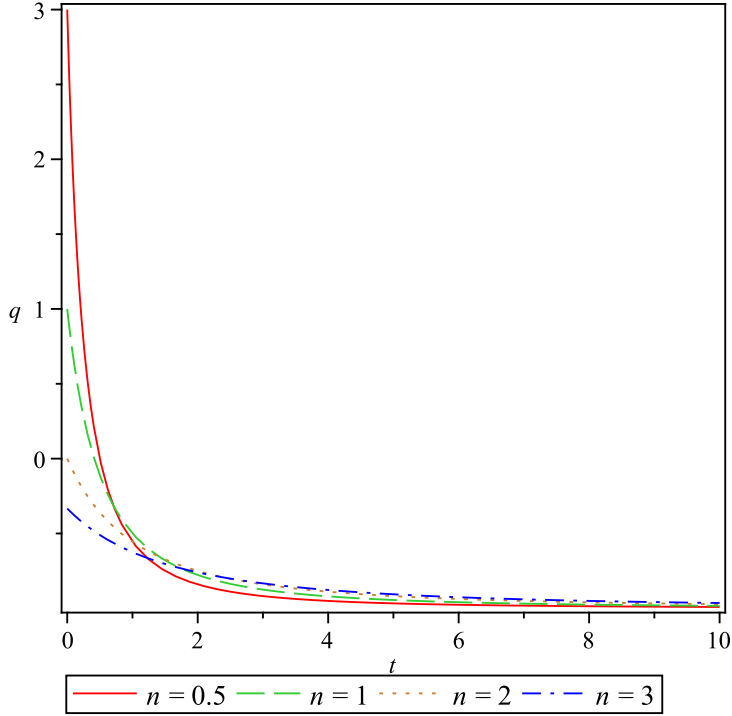} \\
\caption{The deceleration parameter $q$ versus $t$}. 
\end{figure}
Using (\ref{eq31}) into (\ref{eq27})-(\ref{eq29}), we get the following expressions for scale factors:
\begin{equation}
\label{eq32} A(t) = (t^{n}e^{t})^{\frac{1}{2}},
\end{equation}
\begin{equation}
\label{eq33} B(t) = m (t^{n}e^{t})^{\frac{1}{2}}\exp{\left(\ell \int{(t^{n}e^{t})^{-\frac{3}{2}}dt}\right)},
\end{equation}
\begin{equation}
\label{eq34} C(t) = m^{-1}(t^{n}e^{t})^{\frac{1}{2}} \exp{\left(-\ell \int{(t^{n}e^{t})^{-\frac{3}{2}}dt}\right)}.
\end{equation}
The expressions for physical parameters such as directional Hubble parameters ($H_{x}$, $H_{y}$, $H_{z}$), the Hubble 
parameter ($H$), scalar of expansion ($\theta$), shear scalar ($\sigma$), spatial volume $V$ and the anisotropy parameter 
($A_{m}$) are, respectively, given by
\begin{equation}
\label{eq35} H_{x} = \frac{1}{2}\left(\frac{n}{t} + 1\right),
\end{equation}
\begin{equation}
\label{eq36} H_{y} = \frac{1}{2}\left(\frac{n}{t} + 1\right) + \ell \left(t^{n}e^{t}\right)^{-\frac{3}{2}},
\end{equation}
\begin{equation}
\label{eq37}  H_{z} = \frac{1}{2}\left(\frac{n}{t} + 1\right) - \ell \left(t^{n}e^{t}\right)^{-\frac{3}{2}},
\end{equation}
\begin{equation}
\label{eq38} \theta = 3H = \frac{3}{2}\left(\frac{n}{t} + 1\right),
\end{equation}
\begin{equation}
\label{eq39} \sigma^{2} = \ell^{2}\left(t^{n}e^{t}\right)^{-3},
\end{equation}
\begin{equation}
\label{eq40} V = \left(t^{n}e^{t}\right)^{\frac{3}{2}}\exp{(2\alpha x)},
\end{equation}
\begin{equation}
\label{eq41} A_{m} = \frac{8\ell^{2}}{3}\left(\frac{n}{t} + 1\right)^{-2}\left(t^{n}e^{t}\right)^{-3}.
\end{equation}
It is observed that at $t = 0$, the spatial volume vanishes and other parameters $\theta$, $\sigma$, $H$ diverge. 
Hence the model starts with a big bang singularity at $t = 0$. This is a Point Type singularity (MacCallum 1971) 
since directional scale factor $A(t)$, $B(t)$ and $C(t)$ vanish at initial time. \\\\
Figure $1$ depicts the variation of anisotropic parameter ($A_{m}$) versus cosmic time $t$. From the figure, we observe  
that $A_{m}$ decreases with time and tends to zero as $t \to \infty$ for all values of $n$. Thus, the observed isotropy 
of the universe can be achieved in our derived model at present epoch. The shear tensor also tends to zero in this model. \\\\

\begin{figure}[ht]
\centering
\includegraphics[width=8cm,height=8cm,angle=0]{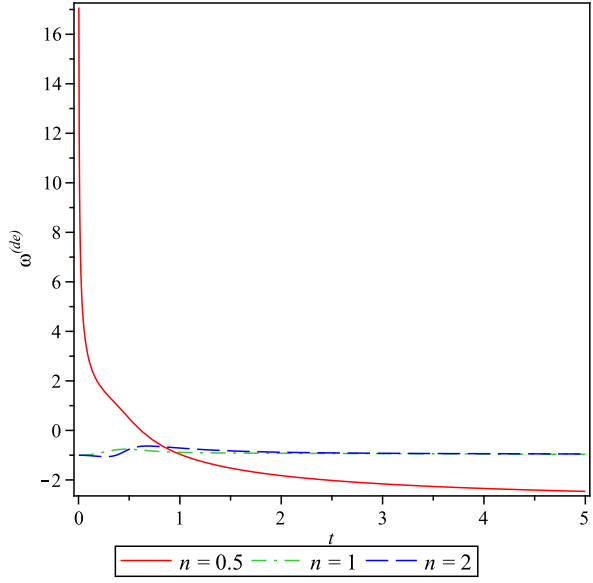} \\
\caption{The plot of DE EoS parameter $\omega^{(de)}$ versus $t$.
Here $\rho_{0} = 1$, $\alpha = 1$, $\ell = 1$, $\omega^{(m)} = 0.5$ }.
\end{figure}
\begin{figure}[ht]
\centering
\includegraphics[width=10cm,height=10cm,angle=0]{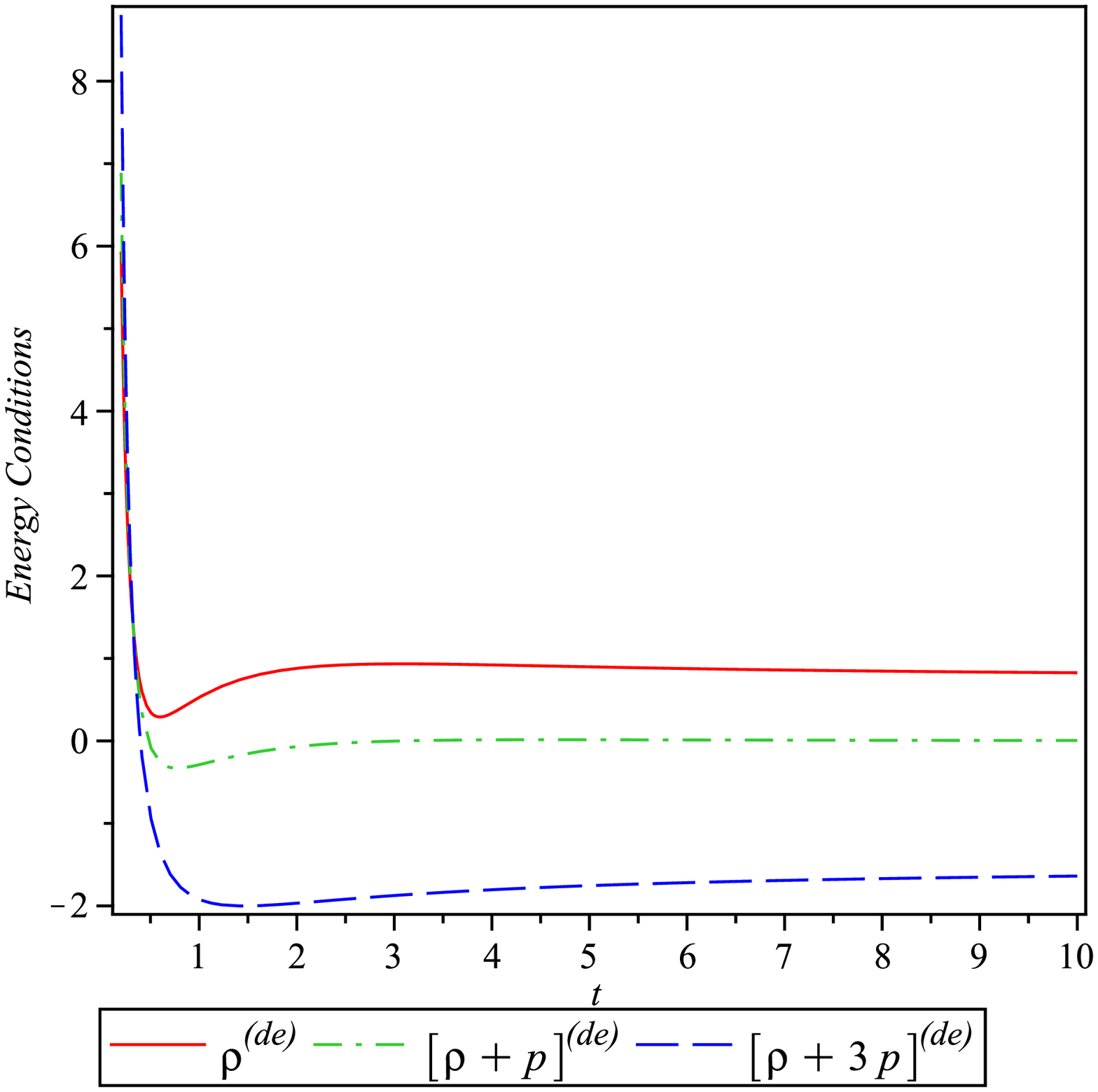} \\
\caption{The plot of energy conditions versus $t$ for $n = 0.5$. 
Here  $\rho_{0} = 1$, $\ell = \alpha = 1$, $\omega^{(m)} = 0.5$}.
\end{figure}
\begin{figure}[ht]
\centering
\includegraphics[width=10cm,height=10cm,angle=0]{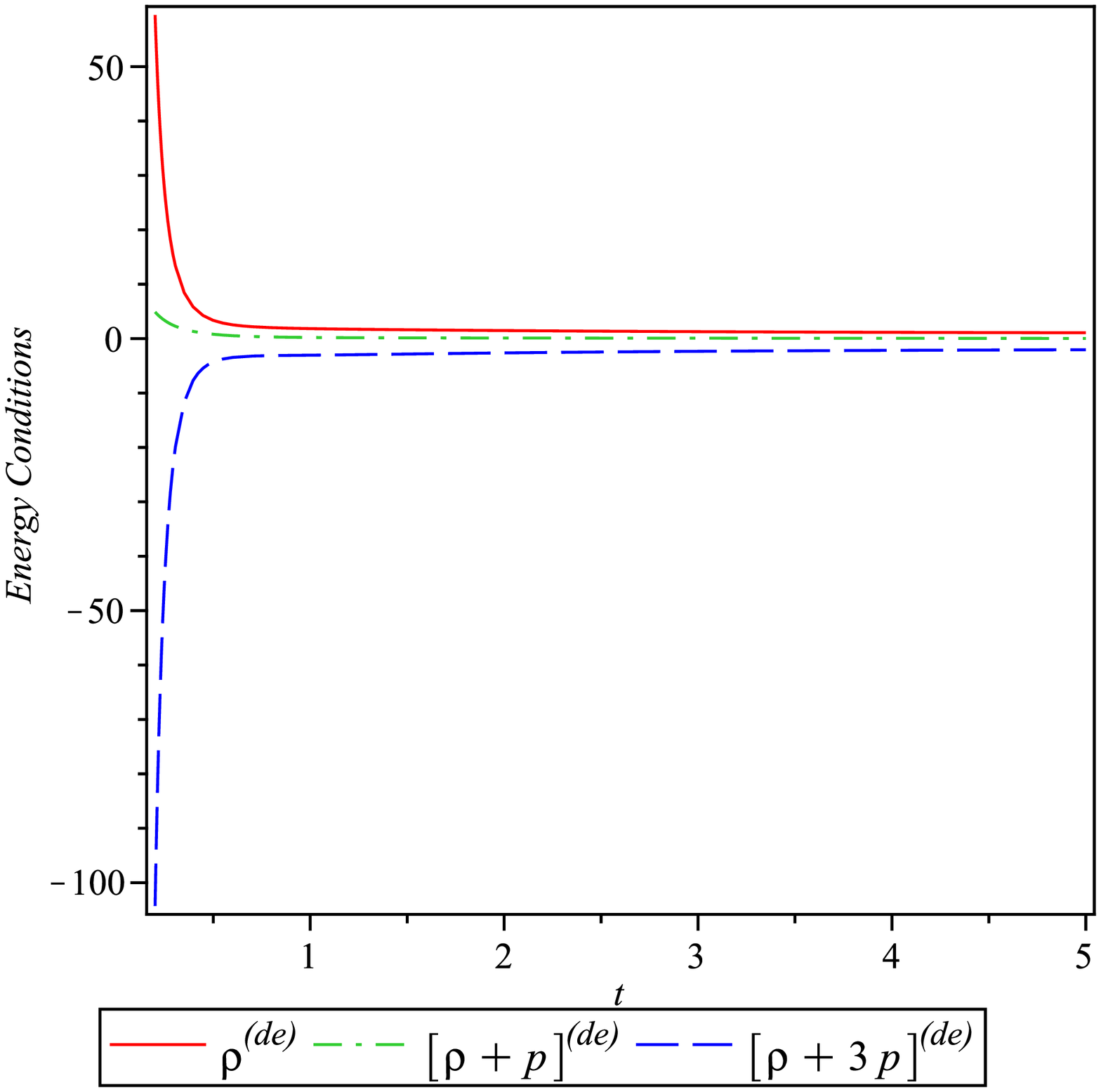} \\
\caption{The plot of energy conditions versus $t$ for $n = 1$. 
Here $\rho_{0} = 1$, $\ell = \alpha = 1$, $\omega^{(m)} = 0.5$}.
\end{figure}
\begin{figure}[ht]
\centering
\includegraphics[width=10cm,height=10cm,angle=0]{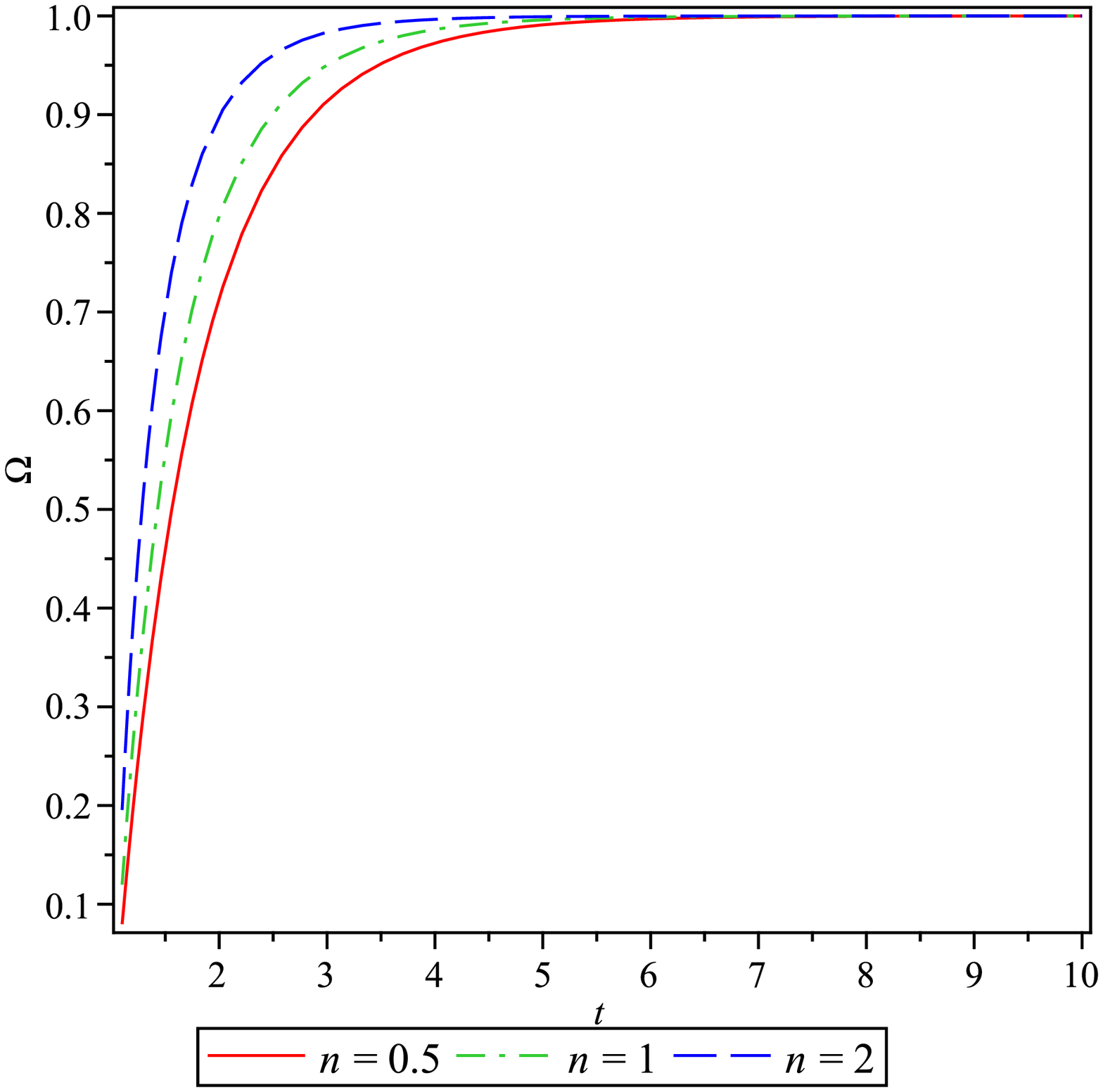} \\
\caption{The total energy density parameter $\Omega$ versus $t$. 
Here $\alpha = \ell = 1$}.
\end{figure}
We define the deceleration parameter $q$ as usual, i.e.
\begin{equation}
\label{eq42} q = - \frac{\ddot{a} a}{\dot{a}^{2}} = - \frac{\ddot{a}}{aH^{2}}.
\end{equation}
Using Eq. (\ref{eq31}) into Eq. (\ref{eq42}), we find
\begin{equation}
\label{eq43}q = \frac{2n}{(n + t)^{2}} - 1.
\end{equation}
From Eq. (\ref{eq43}), we observe that $q > 0$ for $t < \sqrt{2n} - n$ and $q < 0$ for $t > \sqrt{2n} -n$. 
It is observed that for $0 < n < 2$, our model is evolving from deceleration phase to acceleration phase. Also, recent 
observations of SNe Ia, expose that the present universe is accelerating and the value of DP lies to some place in the 
range $-1 < q < 0$. It follows that in our derived model, one can choose the value of DP consistent with the observation. 
Figure $2$ graphs the deceleration parameter ($q$) versus time which gives the behaviour of $q$ from decelerating to 
accelerating phase for different values of $n$. \\\\
The energy density ($\rho^{(m)}$) of perfect fluid, the pressure ($p^{(de)}$) of DE component, DE density ($\rho^{(de)}$) 
and EoS parameter ($\omega^{(de)}$) of DE, for this model are given by 
\begin{equation}
\label{eq44} \rho^{(m)} = \rho_{0}(t^{n}e^{t})^{-\frac{3}{2}(1 + \omega^{(m)})},
\end{equation}
\begin{equation}
\label{eq45} p^{(de)} = -\frac{3}{4}\left(\frac{n}{t} + 1\right)^{2} + \frac{n}{t^{2}} - \ell(t^{n}e^{t})^{-3} 
+ \alpha^{2}(t^{n}e^{t})^{-1} - \omega^{(m)}\rho_{0}(t^{n}e^{t})^{-\frac{3}{2}(1 + \omega^{(m)})},
\end{equation}
\begin{equation}
\label{eq46} \rho^{(de)} = - \rho_{0}(t^{n}e^{t})^{-\frac{3}{2}(1 + \omega^{(m)})} + \frac{3}{4}\left(\frac{n}{t} + 1\right)
^{2} + \ell(t^{n}e^{t})^{-3} - 3\alpha^{2}(t^{n}e^{t})^{-1},
\end{equation}
\begin{equation}
\label{eq47} \omega^{(de)} = - \frac{\frac{3}{4}\left(\frac{n}{t} + 1\right)^{2} - \frac{n}{t^{2}} + \ell(t^{n}e^{t})^{-3} 
- \alpha^{2}(t^{n}e^{t})^{-1} + \omega^{(m)}\rho_{0}(t^{n}e^{t})^{-\frac{3}{2}(1 + \omega^{(m)})}}{\frac{3}{4}\left(\frac{n}
{t} + 1\right)^{2} + \ell(t^{n}e^{t})^{-3} - 3\alpha^{2}(t^{n}e^{t})^{-1} - \rho_{0}(t^{n}
e^{t})^{-\frac{3}{2}(1 + \omega^{(m)})}}.
\end{equation}
Figure $3$ depicts the variation of DE EoS parameter $\omega^{(de)}$ versus cosmic time t. We observe from the figure 
that for $n <1$,  $\omega^{(de)}$ varies from non-dark region crossing the PDL ($\omega^{(de)} = -1$) and ultimately 
approaches to the phantom region ($\omega^{(de)} < -1$). But for $n \geq 1$, the variation of $\omega^{(de)}$ starts 
from cosmological constant region ($\omega^{(de)} = -1$) and finally approached to the quintessence region 
($\omega^{(de)} > -1$). Therefore, we observe that for $n \geq 1$, the variation of $\omega^{(de)}$ in our derived model
is consistent with the recent observations of SNe Ia data \cite{ref14}, SNe Ia data with CMBR anisotropy and 
galaxy clustering statistics \cite{ref6}. \\\\
The dark energy with $\omega^{(de)} < -1$, the phantom component of the universe, leads to uncommon cosmological scenarios 
as it was pointed out by Caldwell et al. \cite{ref42}. First of all, there is a violation of the dominant energy condition 
(DEC), since $\rho + p < 0$. The energy density grows up to infinity in a finite time, which leads to a big rip,
characterized by a scale factor blowing up in this finite time. These sudden future singularities are, nevertheless, 
not necessarily produced by a fluid violating DEC. Cosmological solutions for phantom matter which violates the weak 
energy condition were found by Dabrowski et al. \cite{ref43}. Caldwell \cite{ref44}, Srivastava \cite{ref45},
Yadav \cite{ref24} have investigated phantom models with $\omega^{(de)} < -1$ and also suggested that at late time, phantom 
energy has appeared as a potential DE candidate which violets the weak as well as strong energy condition. \\\\
The left hand side of energy conditions have been depicted in Figures $4$ and $5$ for different values of $n$. 
From Figure $4$, for $n = 0.5$ (i.e phantom model) (see, Figure $3$) , we observe that
$$
(i) ~ ~ \rho^{(de)} \geq 0,  ~ ~ ~ (ii) ~ ~  \rho^{(de)} + p^{(de)} \leq 0,  ~ ~ ~   (iii) ~ ~ \rho^{(de)} + 3p^{(de)} < 0. 
$$
Thus, from above expressions, we observe the phantom model violates both the strong and weak energy conditions, 
as expected.\\\\
Furthe, from Figure $5$, for $n \geq 1$ (i.e quintessence model) (see, Figure $3$), we observe that 
$$
(i) ~ ~ \rho^{(de)} \geq 0,  ~ ~ ~ (ii) ~ ~  \rho^{(de)} + p^{(de)} \geq 0,  ~ ~ ~   (iii) ~ ~ \rho^{(de)} + 3p^{(de)} < 0. 
$$  
Thus, the quintessence model violates the strong energy condition as the same is predicted by current astronomical 
observations. \\\\ 
The perfect fluid density parameter ($\Omega^{(m)}$) and DE density parameter ($\Omega^{(de)}$) are given by 
\begin{equation}
\label{eq48}\Omega^{(m)} = \frac{4}{3}\rho_{0}\left(\frac{n}{t} + 1\right)^{2}(t^{n}
e^{t})^{-\frac{3}{2}(1 + \omega^{(m)})}. 
\end{equation}
\begin{equation}
\label{eq49} \Omega^{(de)} = 1 + \frac{4}{3}\left(\frac{n}{t} + 1\right)^{-2}\left[- \rho_{0} (t^{n}
e^{t})^{-\frac{3}{2}(1 + \omega^{(m)})} + \ell^{2}(t^{n}e^{t})^{-3} - 3\alpha(t^{n}e^{t})^{-1}\right].
\end{equation}
Thus the over all density parameter ($\Omega$) is obtained as 
\[
 \Omega = \Omega^{(m)} + \Omega^{(de)}
\]
\begin{equation}
\label{eq50} = 1 + \ell^{2}(t^{n}e^{t})^{-3} - 3\alpha(t^{n}e^{t})^{-1}.  
\end{equation}
Figure $6$ depicts the variation of the density parameter ($\Omega$) versus cosmic time $t$ for different values of $n$ 
during the evolution of the universe. From the Figure $6$, it can be seen that the total energy density $\Omega$ tends 
to $1$ for sufficiently large time which is supported by the current observations. \\

\section{Cosmic Jerk Parameter}
A convenient method to describe models close to $\Lambda$ CDM is based on the cosmic jerk parameter $j$, a
dimensionless third derivative of the scale factor with respect to the cosmic time \cite{ref46}$-$\cite{ref50}. 
A deceleration-to-acceleration transition occurs for models with a positive value of $j_{0}$ and negative $q_{0}$. 
Flat $\Lambda$ CDM models have a constant jerk $j = 1$. The jerk parameter in cosmology is defined as the dimensionless 
third derivative of the scale factor with respect to cosmic time
\begin{equation}
\label{eq51} j(t) = \frac{1}{H^{3}}\frac{\dot{\ddot{a}}}{a}.
\end{equation}
and in terms of the scale factor to cosmic time
\begin{equation}
\label{eq52} j(t) = \frac{(a^{2}H^{2})^{''}}{2H^{2}}.
\end{equation}
where the `dots' and `primes' denote derivatives with respect to cosmic time and scale factor, respectively.
One can rewrite Eq. (\ref{eq51}) as
\begin{equation}
\label{eq53} j(t) = q + 2q^{2} - \frac{\dot{q}}{H}.
\end{equation}
Eqs. (\ref{eq43}) and (\ref{eq53}) reduce to
\begin{equation}
\label{eq54} j(t) = 1 - \frac{6n}{(n + t)^{2}} + \frac{8n}{(n + 1)^{3}}.
\end{equation}
This value overlaps with the value $j\simeq2.16$ obtained from the combination of three kinematical data sets: the
gold sample of type Ia supernovae \cite{ref9}, the SNIa data from the SNLS project \cite{ref10}, and 
the X-ray galaxy cluster distance measurements \cite{ref51} for
\begin{equation}
\label{eq55} t = 3.45\times10^{-2}A - \frac{50n}{A} - n,
\end{equation}
where $A = 10^{4}n[8.41 + 1.45\sqrt{(14.4n + 33.6)}]^{\frac{1}{3}}$.  
\section{Discussion and Concluding Remarks}
In this paper, we have studied a spatially homogeneous and anisotropic Bianchi type-V space time filled with perfect fluid 
and anisotropic DE possessing dynamic energy density. The field equations have been solved exactly with suitable physical 
assumptions. The solutions satisfy the energy conservation Eq. (\ref{eq17}) and the Raychaudhuri Eq. (\ref{eq18}) 
identically. Therefore, exact and physically viable Bianchi type-V model has been obtained. It is to be noted that our 
procedure of solving the field equations is altogether different from what Kumar \& Yadav \cite{ref25} have adopted. Kumar 
\& Yadav \cite{ref25} have solved the field equations by considering the constant DP whereas we have considered time dependent 
DP. As we have already discussed in previous Sect. $3$ that for a Universe which was decelerating in past and accelerating 
at the present time, the DP must show signature flipping (see the Refs. Padmanabhan and Roychowdhury \cite{ref38}, 
Amendola \cite{ref39}, Riess et al. \cite{ref40} and so, there is no scope for a constant DP. The main features of 
the model are as follows: \\

$\bullet$ For different values of $n$ the anisotropic parameter $A_{m}$ tends to zero for sufficiently large time. Hence, 
the present model is isotropic at late time which is consistent to the current observations.\\

$\bullet$ The present DE model has a transition of the universe from the early deceleration phase to the recent acceleration 
phase (see, Figure $2$) which is in good agreement with recent observations \cite{ref52}. \\

$\bullet$ In the present study we find that for $n \geq 1$, the quintessence model is consistent with present and expected 
future evolution of the universe. The quintessence model approaches to isotropy at late time (see, Figure $1$ \& $3$). 
The other models (for $n < 1$), we observe the phantom scenario. \\

$\bullet$ The derived phantom  model violates both the strong  and weak energy conditions whereas the quintessence model 
violates only the strong energy condition (see, Figures $4$ \& $5$). \\

$\bullet$ The total density parameter ($\Omega$) approaches to $1$ for sufficiently large time (see, Figure $6$) which 
is reproducible with current observations. \\ 

$\bullet$ The cosmic jerk parameter in our descended model is also found to be in good agreement with 
the recent data of astrophysical observations namely the gold sample of type Ia supernovae \cite{ref9}, the 
SNIa data from the SNLS project \cite{ref10}, and the X-ray galaxy cluster distance measurements 
\cite{ref51}. \\

$\bullet$  Our special choice of scale factor yields a time dependent deceleration parameter which represents a model 
of the Universe which evolves from decelerating phase to an accelerating phase whereas in Yadav \cite{ref24}, 
Kumar \& Yadav \cite{ref25} only the evolution takes place either in an accelerating or a decelerating phase. \\

$\bullet$ For different choice of $n$, we can generate a class of DE models in Bianchi type-V space-time. It is observed 
that such DE models are also in good harmony with current observations. Thus, the solutions demonstrated in this paper may 
be useful for better understanding of the characteristic of anisotropic DE in the evolution of the universe within the 
framework of Bianchi type-V space-time.
\section*{Acknowledgments}
Author (AP) would like to thank the Inter-University Centre for Astronomy and Astrophysics (IUCAA), Pune, India 
for providing facility and support where part of this work was carried out. 

\end{document}